# Slow-roll inflation in $f(R, T, R_{ab}T^{ab})$ gravity

Zhe Feng
*College of Science, Hohai University,*
*Nanjing 211100, P. R. China*
*2010020129@hhu.edu.cn*

Received 15 September 2023
Revised 22 January 2024
Accepted 16 February 2024
Published 8 April 2024

In the framework of $f(R, T, R_{ab}T^{ab})$ gravity theory, the slow-roll approximation of the cosmic inflation is investigated, where $T$ is the trace of the energy–momentum tensor $T^{ab}$, $R$ and $R_{ab}$ are the Ricci scalar and tensor, respectively. After obtaining the equations of motion of the gravitational field from the action principle in the spatially flat FLRW metric, the fundamental equations of this theory are received by introducing the inflation scalar field as the matter and taking into account only the minimum curvature-inflation coupling term. Remarkably, after taking the slow-roll approximation, the identical equations as in $f(R, T)$ gravity with a $RT$ mixing term are derived. We study several potentials of interest in different domains. We perform analytical analyzes under various approximate conditions, and present numerical results and their comparison with existing observational data at the same time. In the appendix, we analyze the behavior of the inflation scalar field under perturbation while ignoring the effect of metric perturbations. This research complements the slow-roll inflation in the modified theory of gravity.

## 1. Introduction

As a concise and elegant theory of gravity, general relativity (GR) has achieved significant success in the field of cosmology. $\Lambda$CDM theory in the framework of GR is regarded as the standard model of cosmology and consistent with numerous experimental results in the past decades.[1] However, some traditional cosmological issues, as well as the more refined observational problems brought about by the continuous progress of technological tools, especially the flatness problem and the horizon problem, were unable to find convincing answers within the preexisting framework.[2] A few decades ago, some authors[3–6] developed techniques to try to address these problems by considering the inclusion of an exponential expansion phase in the early universe, called inflation. The most straightforward and popular method is to introduce a scalar field and consider the behavior of inflation at different potentials.[7] By imposing the slow-roll approximation, i.e. by appropriately neglecting the higher-order terms, it is usually possible to obtain succinct equations of motion. This idea is being examined.[8,9]

The GR has the smallest curvature-matter coupling as observed from the action perspective. The deviation of experimental observations from the theory has led to an interest in the dark sector, including dark matter and dark energy. Einstein's cosmological constant can be considered as a pioneering attempt, and after that, the search for more complex possibilities has since become one of the main motivations to study various theories of gravity.[10,11] The $f(R)$ gravity theory[12,13] is a simple yet effective effort to achieve this by substituting the Ricci scalar in the action under GR with an arbitrary function of it. This model did not enhance the coupling between curvature and matter, which was subsequently remedied by the $f(R,T)$ gravity theory.[14–16] About a decade ago, $f(R,T,R_{ab}T^{ab})$ gravity theory[17,18] was proposed, and this model apparently contains a stronger coupling between curvature and matter, further deepening our grasp of spacetime. In contrast to GR, when there is a non-minimal gravitational-matter coupling, the standard conservation of energy–momentum fails and is replaced by an equation containing a curvature part,[19] which can be obtained by finding the covariance derivative of the field equation. This implies that there is an energy–momentum exchange between the matter and gravitational fields, which may give rise to interesting phenomena in the investigation of quantum gravity.

The slow-roll inflation in the $f(R)$ and $f(R,T)$ gravitational frames has been research done.[20,21] Recently, some authors[22] explored the case in $f(R,T)$ gravity with non-minimal coupling, i.e. with a $RT$ mixing term. Inspired by it, we carry out the present work to examines a similar case under $f(R,T,R_{ab}T^{ab})$ gravity. Starting from the action principle, after obtaining the equations of motion for the gravitational field and the inflation scalar field separately and taking the slow-roll approximation, we can obtain the same set of equations as in Ref. 22. This implies that we can study more intricate physical situations without increasing the mathematical difficulty. Various inflation models have been proposed for different problems, which can be distinguished according to different potential functions, a few typical examples such as Refs. 23–28. We provide a brief introduction to the potentials under consideration and calculate their corresponding slow-roll parameters and e-folding number $N$ under a series of approximate conditions, which is of great benefit to our understanding of the behavior of inflationary dynamics. We then give numerical results of observables $n_s$ and $r$ corresponding to several potential functions and provide scatter plots with trend descriptions. In the appendix, we provide some calculation details and consider a simplified version of the cosmological perturbation theory, which ignores the scalar and tensor perturbations on the metric and only considers a perturbation that varies with time and space on the spatially homogeneous and isotropic background scalar field, the dynamic equation of this disturbance field is obtained.

The paper is organized as follows. In Sec. 2, we review the basic framework of $f(R,T,R_{ab}T^{ab})$ gravity theory, introduce the inflation scalar field, and settle on a particular $f(R,T,R_{ab}T^{ab})$ functional form. In Sec. 3, we introduce the slow-roll approximation and derive the equations of the inflation model under this approximation. We also carried out a transformation here from the Jordan to the Einstein framework, which led to the definition of the slow-roll parameter. In Sec. 4, we consider several potential functions of interest and give some basic computational results. In Sec. 5, we present numerical results predicted by theory and compare them with current observations. Some clarifications on the numerical results are provided. As a conclusion, Sec. 6 makes a summary. Some computational details are provided in Appendix A. The behavior of the inflation scalar field under perturbations is taken into consideration in Appendix B.

## 2. Framework of $f(R,T,R_{ab}T^{ab})$ Gravity Theory

We start with the general action of $f(R,T,R_{ab}T^{ab})$ gravity[17,18]

$$S = S_G + S_m = \int \mathrm{d}^4x\sqrt{-g}\,\frac{f(R,T,R_{ab}T^{ab})}{2\kappa} + \int \mathrm{d}^4x\sqrt{-g}\mathcal{L}_m, \tag{1}$$

where $\kappa \equiv 8\pi G \equiv 1/M_{\mathrm{Pl}}^2$ with $G$ being gravitational constant and $g$ is the determinant of the metric $g_{ab}$. $f$ is an arbitrary function of the Ricci scalar $R$, the trace of the energy–momentum tensor $T$ and the coupling term. $\mathcal{L}_m$ is the matter Lagrangian. We predetermine that the matter field is a inflation scalar field in order to give some computational results:

$$\mathcal{L}_m = -\frac{1}{2}g^{ab}\nabla_a\phi\nabla_b\phi - V(\phi). \tag{2}$$

Then the energy–momentum tensor can be defined as the variational derivative of $\mathcal{L}_m$,

$$T_{ab} = \frac{-2}{\sqrt{-g}}\frac{\delta(\sqrt{-g}\mathcal{L}_m)}{\delta g^{ab}} = g_{ab}\mathcal{L}_m - 2\frac{\partial\mathcal{L}_m}{\partial g^{ab}} = \partial_a\phi\partial_b\phi + g_{ab}\mathcal{L}_m. \tag{3}$$

According to the action principle, equations of motion (EoM) of gravitational field are obtained by varying the action Eq. (1) with respect to the metric $g^{ab}$,

$$\begin{aligned}\delta S = {}& \frac{1}{2\kappa}\int \mathrm{d}^4x[\delta\sqrt{-g}f(R,T,R_{ab}T^{ab}) + \sqrt{-g}(f_R\delta R + f_T\delta T + f_P\delta(R_{ab}T^{ab}))] \\ & + \int \mathrm{d}^4x\delta(\sqrt{-g}\mathcal{L}_m),\end{aligned} \tag{4}$$

$R_{ab}T^{ab}$ is written as $P$ and the subscript of $f$ means derivative. The procedures for attributing each term to the variation of the metric $\delta g^{ab}$ are as follows:

$$\delta(\sqrt{-g}) = -\frac{1}{2}\sqrt{-g}\;g_{ab}\delta g^{ab}, \tag{5}$$

$$f_R\delta R + f_T\delta T = [R_{ab}f_R + (g_{ab}\Lambda - \nabla_a\nabla_b)f_R + (T_{ab} + \Theta_{ab})f_T]\delta g^{ab}, \tag{6}$$

$$f_P\delta(R_{ab}T^{ab}) = f_P(T^{ab}\delta R_{ab} + R_{ab}\delta T^{ab}), \tag{7}$$

$$f_P T^{ab}\delta R_{ab} = \left[\frac{1}{2}\Lambda(f_P T_{ab}) + \frac{1}{2}g_{ab}\nabla_c\nabla_d(f_P T^{cd}) - \nabla_c\nabla_b(f_P g^{cd}T_{ad})\right]\delta g^{ab}, \tag{8}$$

$$f_P R_{ab}\delta T^{ab} = f_P\left[-G_{ab}\mathcal{L}_m - \frac{1}{2}RT_{ab} + 2g^{cd}R_{da}T_{bc} - 2R^{cd}\frac{\delta^2\mathcal{L}_m}{\delta g^{ab}\delta g^{cd}}\right]\delta g^{ab} \tag{9}$$

$$\equiv f_P\Xi_{ab}\delta g^{ab}. \tag{10}$$

We define two auxiliary tensors in this process:

$$\Theta_{ab} \equiv g^{cd}\frac{\delta T_{cd}}{\delta g^{ab}} = g_{ab}\mathcal{L}_m - 2T_{ab} - 2g^{cd}\frac{\delta\mathcal{L}_m}{\delta g^{ab}\delta g^{cd}} = -\partial_a\phi\partial_b\phi - T_{ab}, \tag{11}$$

$$\Xi_{ab} \equiv -G_{ab}\mathcal{L}_m - \frac{1}{2}RT_{ab} + 2g^{cd}R_{d(a}T_{b)c}. \tag{12}$$

It should be emphasized that, due to the symmetry of metric $g_{ab}$, there are brackets in the equation that indicate symmetrization. The general form of the gravitational field equation can be produced by summarizing the aforementioned terms:

$$R_{ab}f_R - \frac{1}{2}g_{ab}f + (g_{ab}\Lambda - \nabla_a\nabla_b)f_R + (T_{ab}+\Theta_{ab})f_T + \frac{1}{2}\Lambda(T_{ab}f_P)$$
$$+ \frac{1}{2}g_{ab}\nabla_c\nabla_d(T^{cd}f_P) - \nabla_c\nabla_{(a}(g^{cd}T_{b)d}f_P) + \Xi_{ab}f_P = \kappa T_{ab}. \tag{13}$$

The modified energy–momentum conservation equation can be derived by finding the covariance derivative of the field equation. We will derive it below for specific problems, but do not supply the generic form here.

In the rest part of this paper, we will just consider the following functional form of $f(R,T,R_{ab}T^{ab})$, which is actually the minimal (linear) coupling form in this framework:

$$f(R,T,R_{ab}T^{ab}) = R(1+\alpha) + \gamma\kappa T + 4\beta\kappa^2 R_{ab}T^{ab},\quad f_R = 1+\alpha,\quad f_T = \gamma\kappa,\quad f_P = 4\beta\kappa^2, \tag{14}$$

where $\alpha$, $\gamma$ and $\beta$ are dimensionless constant. Concomitantly, the field equations can be further simplified:

$$R_{ab}(1+\alpha) - \frac{1}{2}g_{ab}f + \gamma\kappa(T_{ab}+\Theta_{ab})$$
$$+ 4\beta\kappa^2\left[\frac{1}{2}\Lambda T_{ab} + \frac{1}{2}g_{ab}\nabla_c\nabla_d T^{cd} - \nabla_c\nabla_{(a}(g^{cd}T_{b)d}) + \Xi_{ab}\right] = \kappa T_{ab}. \tag{15}$$

Since we haven't yet qualified the form of the metric gauge, our analysis is applicable to a spatially non-flat universe. However, since observations seem to indicate that the universe is approximately flat on large scales, we only consider the spatially flat Friedmann–Lemaître–Robertson–Walker (FLRW) metric

$$\mathrm{d}^2s = -\mathrm{d}^2t + a(t)^2(\mathrm{d}^2x + \mathrm{d}^2y + \mathrm{d}^2z), \tag{16}$$

where $t$ is the cosmic time and $a(t)$ is the scale factor. Usually define the Hubble function as $H \equiv \dot{a}/a$ with $\dot{}$ meaning the dervatives with respect to $t$. Details of the computation of some tensor under this metric can be found in Appendix A. The two independent components of the EoMs are as follows:

$$H^2(1+\alpha) = \frac{\kappa}{3}\left[\frac{\dot{\phi}^2}{2}(1+\gamma+18\beta\kappa H^2) + V(\phi)(1+2\gamma+12\beta\kappa H^2)\right] + 4\beta\kappa^2 H\dot{\phi}V'(\phi), \tag{17}$$

$$\dot{H}[1+\alpha-4\beta\kappa^2 V(\phi)] = -\kappa\frac{\dot{\phi}^2}{2}[1+\gamma+12\beta\kappa H^2 - 4\beta\kappa\dot{H}] + 2\beta\kappa^2[\dot{\phi}^2 V''(\phi) + \ddot{\phi}V'(\phi) + 2H\dot{\phi}\ddot{\phi} - H\dot{\phi}V'(\phi)]. \tag{18}$$

The equations of motion of the scalar field, i.e. the modified Klein–Gordon equation, can be obtained by finding the time derivatives of the gravitational field equations (17) and (18) and substituting each other, or by finding the covariant derivatives of the general form of the gravitational field equations (13) or by varying the action Eq. (1) with respect to the scalar field:

$$\ddot{\phi}[1+\gamma+12\beta\kappa H^2] + 3H\dot{\phi}[1+\gamma+12\beta\kappa H^2 + 8\beta\kappa\dot{H}] + V'(\phi)[1+2\gamma+24\beta\kappa H^2 + 18\beta\kappa\dot{H}] = 0. \tag{19}$$

Equations (17)–(19) are consistent with Ref. 21 when $\alpha = \beta = 0$ and naturally return to the case of GR when $\alpha = \gamma = \beta = 0$. They will serve as the fundamental foundation for the following research.

## 3. Slow-Roll Inflation

Considering the time-dependent quantities for quasi-static evolution, the following approximation can be taken:

$$|\dot{\phi}^2| \ll |V(\phi)|, \quad |\ddot{\phi}| \ll |H\dot{\phi}|, \quad |\ddot{H}| \ll |H\dot{H}| \ll |H^3|. \tag{20}$$

Immediately thereafter, Eqs. (17)–(19) can be approximated as

$$H^2(1+\alpha) - \frac{\kappa}{3}V(\phi)(1+2\gamma+12\beta\kappa H^2) = 0, \tag{21}$$

$$\dot{H}(1+\alpha-4\beta\kappa^2 V(\phi)) + \kappa\frac{\dot{\phi}^2}{2}(1+\gamma+12\beta\kappa H^2) + 2\beta\kappa^2 H\dot{\phi}V'(\phi) = 0, \tag{22}$$

$$3H\dot{\phi}(1+\gamma+12\beta\kappa H^2) + V'(\phi)(1+2\gamma+24\beta\kappa H^2) = 0. \tag{23}$$

We are surprised to find that this set of equations has the same results as in Ref. 22. Also, the non-approximate equations in Ref. 22 have terms with higher order. This can be explained by the fact that $RT$ in Ref. 22 involves the cross terms of the temporal and spatial components of $R_{ab}$ and $T^{ab}$, which are not present in $R_{ab}T^{ab}$ in our action Eq. (1). These cross terms suggest a more intricate interchange of energy–momentum between the matter and gravitational fields, which results in a higher order. This consistency suggests that the analysis performed by Ref. 22 that we are about to extend can be applied equally to the action

$$S = \frac{1}{2\kappa}\int \mathrm{d}^4x\sqrt{-g}[R(1+\alpha) + \gamma\kappa T + 4\beta_1\kappa^2 R_{ab}T^{ab} + \beta_2\kappa^2 RT] + \int \mathrm{d}^4x\sqrt{-g}\mathcal{L}_m, \quad \beta_1 + \beta_2 \equiv \beta. \tag{24}$$

A common step in the study of modified gravity theories is to make the equations under study have the same form as in the standard GR, called the Einstein frame, by means of metric conformal transformations and field redefinitions. This is just a mathematical trick,[29] but can facilitate our finding the observable measurements of great interest to observational cosmology, since their form in GR is familiar. Although in general the strict Einstein frame is not solvable in the presence of non-minimal couplings, the slow-roll approximation helps us to solve this obstacle. The first step in the procedure is to define three helper functions as follows:

$$\tilde{g}_{ab} \equiv \Omega_1(\phi) g_{ab}, \tag{25}$$

$$\text{i.e. } \tilde{g}_{\mu\nu}\mathrm{d}\tilde{x}^\mu \mathrm{d}\tilde{x}^\nu = -\mathrm{d}\tilde{t}^2 + \tilde{a}(\tilde{t})^2 \mathrm{d}x_i \mathrm{d}x^i \equiv \Omega_1(\phi)(-\mathrm{d}t^2 + a(t)^2 \mathrm{d}x_i \mathrm{d}x^i), \tag{26}$$

$$\Omega_2(\phi) \equiv \left(\frac{\mathrm{d}\chi}{\mathrm{d}\phi}\right)^2, \tag{27}$$

$$\Omega_3(\phi) \equiv \frac{\tilde{V}(\chi(\phi))}{V(\phi)}. \tag{28}$$

The expected field equation takes the form

$$3\tilde{H}\frac{\mathrm{d}\chi}{\mathrm{d}\tilde{t}} + \tilde{V}'(\chi) = 0, \tag{29}$$

$$\tilde{H}^2 - \frac{\kappa}{3}\tilde{V}(\chi) = 0, \tag{30}$$

$$\frac{\mathrm{d}\tilde{H}}{\mathrm{d}\tilde{t}} + \frac{\kappa}{2}\left(\frac{\mathrm{d}\chi}{\mathrm{d}\tilde{t}}\right)^2 = 0. \tag{31}$$

These are exactly the slow-roll equations in GR. Substituting Eqs. (26), (27) and (28) into Eqs. (17), (18) and (19) and using care Eqs. (29), (30) and (31), we can invert the solution to Eqs. (26), (27) and (28) as follows:

$$\Omega_1 = 1 + \alpha - 4\beta\kappa^2 V(\phi), \tag{32a}$$

$$\Omega_2 = \frac{(1+\alpha)(1+\gamma) + 4\beta\gamma\kappa^2 V(\phi)}{(1+\alpha-4\beta\kappa^2 V(\phi))^2}, \tag{32b}$$

$$\Omega_3 = \frac{1+2\gamma}{(1+\alpha-4\beta\kappa^2 V(\phi))^2}. \tag{32c}$$

We can look at this set of equations from two perspectives. On the one hand, when the coefficient $\beta$ of the non-minimum coupling term is very small, we can avoid pursuing complex analytical solutions to the specific potential function and instead be satisfied with the approximate transformation relationship

$$\Omega_1 = 1+\alpha+\mathcal{O}(\beta), \tag{33a}$$

$$\Omega_2 = \frac{1+\gamma}{1+\alpha}+\mathcal{O}(\beta), \tag{33b}$$

$$\Omega_3 = \frac{1+2\gamma}{(1+\alpha)^2}+\mathcal{O}(\beta) \tag{33c}$$

and further get

$$\begin{aligned}\chi(t) &= \sqrt{\frac{1+\gamma}{1+\alpha}}\phi(t)+\mathcal{O}(\beta),\\ \tilde{V}(\chi) &= \frac{1+2\gamma}{(1+\alpha)^2}V\left(\sqrt{\frac{1+\alpha}{1+\gamma}}\chi+\mathcal{O}(\beta)\right)+\mathcal{O}(\beta).\end{aligned} \tag{34}$$

In actuality, we can now rescale the scalar field and the potential function to recast the action Eq. (24) as a minimal coupling form of standard GR and a scalar field with a trivial action

$$\tilde{S} = \int \mathrm{d}^4\tilde{x}\sqrt{-\tilde{g}}\left(\frac{\tilde{R}}{2\kappa}-\frac{1}{2}\tilde{g}^{ab}\nabla_a\chi\nabla_b\chi-\tilde{V}(\chi)\right). \tag{35}$$

In other words, $\beta \neq 0$ is necessary in order to understand the impact of $\alpha$ and $\gamma$ on the inflationary observables. On the other hand, since the physics in $f(R,T)$ gravity has long been known, when we consider the influence of the $R_{ab}T^{ab}$ term, we can directly let $\alpha=\gamma=0$ to get

$$\Omega_2=\Omega_3=\Omega_1^{-2}=(1-4\beta\kappa^2V(\phi))^{-2}=1+8\beta\kappa^2V(\phi)+\mathcal{O}(\beta^2). \tag{36}$$

Again, this means formally

$$\mathrm{d}\chi = (1+4\beta\kappa^2V(\phi)+\mathcal{O}(\beta^2))\mathrm{d}\phi \Rightarrow \chi = \phi+4\beta\kappa^2\int V(\phi)\mathrm{d}\phi+\mathcal{O}(\beta^2), \tag{37a}$$

$$\tilde{V}(\chi(\phi)) = (1+8\beta\kappa^2V(\phi)+\mathcal{O}(\beta^2))V(\phi). \tag{37b}$$

In standard GR, considering the last equation in Eq. (20), we can define

$$\epsilon \equiv -\dot{H}/H^2 \ll 1 \quad \text{and} \quad \eta \equiv -\ddot{H}/(2H\dot{H}) \ll 1. \tag{38}$$

The coefficients have been deliberately chosen so that these two quantities satisfy the relationship $\dot{\epsilon} = 2\dot{H}^2/H^3 - \ddot{H}/H^2 = 2H\epsilon(\epsilon-\eta)$. The end of the inflation is understood as the moment when $\epsilon$ or $\eta$ reaches unity. Further using the slow-roll equations, they can be reduced to only potential-dependent form and two equivalent parameters are defined:

$$\epsilon = -\frac{\dot{H}}{H^2} = \frac{1}{2\kappa}\left(\frac{V'(\phi)}{V(\phi)}\right)^2 \equiv \epsilon_V, \tag{39}$$

$$\eta = -\frac{\ddot{H}}{2H\dot{H}} = \frac{1}{\kappa}\left(\frac{V''(\phi)}{V(\phi)}\right) - \frac{1}{2\kappa}\left(\frac{V'(\phi)}{V(\phi)}\right)^2 \equiv \eta_V - \epsilon_V. \tag{40}$$

For the case of $f(R, T, R_{ab}T^{ab})$, we directly define

$$\epsilon_{\tilde{V}} \equiv \frac{1}{2\kappa}\left(\frac{\tilde{V}'(\chi)}{\tilde{V}(\chi)}\right)^2, \quad \eta_{\tilde{V}} \equiv \frac{1}{\kappa}\left(\frac{\tilde{V}''(\chi)}{\tilde{V}(\chi)}\right). \tag{41}$$

In principle, it is still possible to start with the definition equation (38) and obtain parameters solely connected to the potential. However, the process is mathematically complex, and arriving at a usable expression requires a more tedious approximation. Additionally, there are derived quantities that are of interest in light of astronomical observations[30]:

$$n_{\mathrm{S}} \equiv 1 - 6\epsilon_{\tilde{V}} + 2\eta_{\tilde{V}}, \quad n_{\mathrm{T}} \equiv -2\epsilon_{\tilde{V}}, \quad r_* \equiv 16\epsilon_{\tilde{V}}, \tag{42}$$

where $n_{\mathrm{S}}$ and $n_{\mathrm{T}}$ are the scalar and tensor spectral index, respectively, and $r_*$ is the tensor-to-scalar ratio. The e-folding number $N$, defined by $N \equiv \ln[a(t_{\mathrm{end}})/a(t)]$, can be transcribed as

$$N = \int_{\tilde{t}}^{\tilde{t}_{\mathrm{end}}} \tilde{H}(\tilde{t})\mathrm{d}\tilde{t} = \int_{\chi}^{\chi_{\mathrm{end}}} \frac{H}{\mathrm{d}\chi/\mathrm{d}\tilde{t}}\mathrm{d}\chi = \kappa\int_{\chi_{\mathrm{end}}}^{\chi} \frac{\tilde{V}(\chi)}{\tilde{V}'(\chi)}\mathrm{d}\chi, \tag{43}$$

which reflects the degree of spatio-temporal expansion. At this point, our theory has been constructed. The difference between the various models lies in the difference of potential functions. In the next section, we will discuss various forms of potential functions, which are of interest in different domains, and analyze the implications for slow-rolling inflation.

## 4. Inflationary Models: Analytical Results

### 4.1. *Starobinsky inflation*

Alexei Starobinsky from the Soviet Union noted that quantum corrections to GR should be significant for the early universe. These generally result in curvature-squared corrections to the Einstein–Hilbert action and a form of $f(R)$ modified gravity. When the curvatures are enormous, the solution to Einstein's equations with curvature-squared terms yields an effective cosmological constant. Therefore, he suggested an inflationary de Sitter era occurred in the early universe.[31] This resolved the cosmological issues and produced precise estimates for the microwave background radiation corrections, which were later meticulously calculated. Starobinsky originally used the semiclassical Einstein equations with free quantum matter fields.[3] However, it soon became clear that the inflation was mostly managed by the effective action's squared Ricci scalar contribution[32]

$$S=\frac{1}{2\kappa}\int \mathrm{d}^4x\sqrt{-g}\left(R+\frac{R^2}{6M^2}\right), \tag{44}$$

where $M$ is a dimensioned constant. Although this action does not contain a scalar field, in the Einstein framework it corresponds to a coupling of curvature and scalar field[33–35]

$$S=\int \mathrm{d}^4x\sqrt{-g}\left(\frac{R}{2\kappa}-\frac{1}{2}g^{ab}\nabla_a\phi\nabla_b\phi-V(\phi)\right),\quad V(\phi)=\frac{3M^2}{4\kappa}\left[1-\mathrm{e}^{-\sqrt{\frac{2}{3}}\frac{\phi}{M_{\mathrm{Pl}}}}\right]^2 \tag{45}$$

and $M=1.13\times10^{-5}m_{\mathrm{Pl}}$ is required from the scalar fluctuation analysis.[36] This potential with strong asymmetry and distinct asymptotic behavior is shown in Fig. 1. In fact, Higgs inflation with the action

$$S=\int \mathrm{d}^4x\sqrt{-g}\left[-\frac{M^2+\xi h^2}{2}R+\frac{\partial_\mu h\partial^\mu h}{2}-\frac{\lambda}{4}(h^2-v^2)^2\right] \tag{46}$$

can also be reduced to the form of a scalar field coupled to curvature controlled by a potential of this form.[28] The slow roll parameters can be derived by using the already-obtained approaches

$$\epsilon_{\tilde V}=\frac{4}{3}\left[\exp\left(\sqrt{\frac{2}{3}}\frac{\chi}{\widetilde{M_{\mathrm{Pl}}}}\right)-1\right]^2,\quad \eta_{\tilde V}=-\frac{4}{3}\frac{\exp\left(\sqrt{\frac{2}{3}}\frac{\chi}{\widetilde{M_{\mathrm{Pl}}}}\right)-2}{\left[\exp\left(\sqrt{\frac{2}{3}}\frac{\chi}{\widetilde{M_{\mathrm{Pl}}}}\right)-1\right]^2}, \tag{47}$$
$$\widetilde{M_{\mathrm{Pl}}}\equiv\sqrt{\frac{1+\gamma}{1+\alpha}}M_{\mathrm{Pl}}.$$

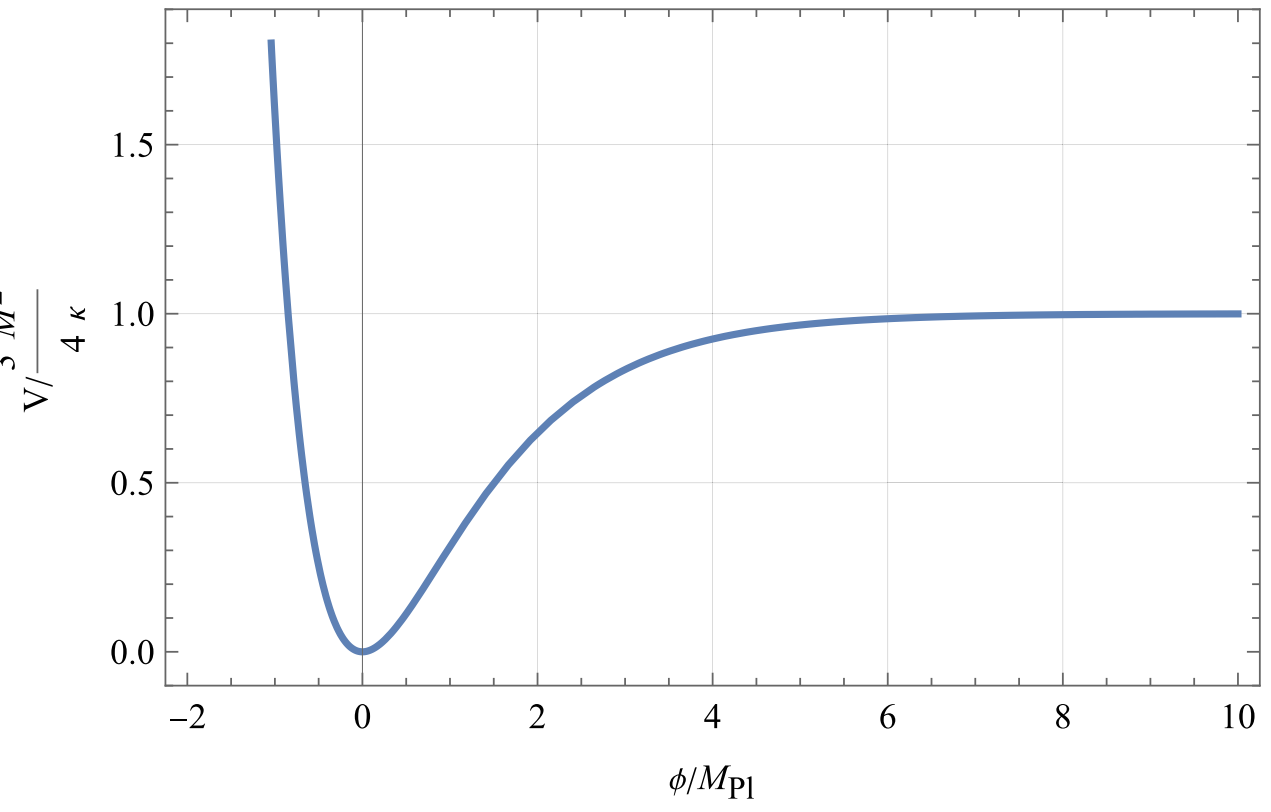

Fig. 1. Starobinsky potential with strong asymmetry and distinct asymptotic behavior.

Further, we can obtain the constraints required by the slow-roll parameter and the e-folding number

$$\epsilon_{\tilde{V}} \leq 1, \quad \eta_{\tilde{V}} \leq 1 \Rightarrow \chi \geq \chi_{\text{end}} \equiv \frac{\sqrt{6}}{2} \ln\left[1 + \frac{2}{\sqrt{3}}\right] \widetilde{M_{\text{Pl}}}, \tag{48}$$

$$N = \frac{1}{4}\left[\left(3e^{\sqrt{\frac{2}{3}}\frac{\chi}{\widetilde{M_{\text{Pl}}}}} - \sqrt{6}\frac{\chi}{\widetilde{M_{\text{Pl}}}}\right) - \left(3 + 2\sqrt{3} - 3\ln\left[1 + \frac{2}{\sqrt{3}}\right]\right)\right]. \tag{49}$$

However, inflation exhibits a slightly more complex tendency under the appropriate initial conditions. For a discussion of the initial conditions of inflation with Starobinsky and more potential, see Ref. 37. If additional estimates are taken into account, one can take

$$N = \frac{1}{4}\left[3e^{\sqrt{\frac{2}{3}}\frac{\chi}{\widetilde{M_{\text{Pl}}}}} - (3 + 2\sqrt{3})\right], \quad \frac{\chi}{\widetilde{M_{\text{Pl}}}} = \sqrt{\frac{3}{2}} \ln\left(\frac{4}{3}N + 1 + \frac{2}{\sqrt{3}}\right). \tag{50}$$

For large $N$,

$$\epsilon_{\tilde{V}} = \frac{12}{(3 - 4N)^2} \approx \frac{3}{4N^2}, \quad \eta_{\tilde{V}} = \frac{8(3 - 2N)}{(3 - 4N)^2} \approx -\frac{1}{N}. \tag{51}$$

This finding is consistent with that in the standard GR[21] because we operate within the Einstein framework and all terms related to $\beta$ are discarded in the small amount approximation.

### 4.2. *Chaotic models with power-law potentials*

Power-law potential has a concise form

$$V(\phi) = \lambda M_{\text{Pl}}^4 \left(\frac{\phi}{M_{\text{Pl}}}\right)^n, \quad n > 0, \tag{52}$$

where $\lambda$, $n$ are dimensionless constants. The case of $n = 4$ was first presented in Ref. 23. The slow-roll parameters are calculated as follows:

$$\epsilon_{\tilde{V}} = \frac{n^2}{2\kappa\chi^2}, \quad \eta_{\tilde{V}} = \frac{n(n-1)}{\kappa\chi^2}. \tag{53}$$

We find that the results are identical to those in the standard GR and do not exhibit the correction generated by $f(R, T, R_{ab}T^{ab})$ gravity. This is because we ignore the higher-order terms about $\beta$ in Eq. (34). Discussing the values of $n$, we get

$$\chi_{\text{end}} = \begin{cases} \dfrac{n}{\sqrt{2\kappa}}, & n \geq 2, \\ \dfrac{\sqrt{n(n-1)}}{\sqrt{\kappa}}, & n < 2, \end{cases} \quad N = \begin{cases} \dfrac{\kappa}{2n}\left(\chi^2 - \dfrac{n^2}{2\kappa}\right), & n \geq 2, \\ \dfrac{\kappa}{2n}\left(\chi^2 - \dfrac{n(n-1)}{\kappa}\right), & n < 2, \end{cases} \tag{54}$$

$$\Rightarrow \chi^2 = \begin{cases} \frac{1}{\kappa}(2nN + n^2/2), & n \geq 2, \\ \frac{1}{\kappa}(2nN + n(n-1)), & n < 2, \end{cases} \tag{55}$$

$$\begin{cases} \epsilon_{\tilde{V}} = \frac{1}{4N/n + 1}, \quad \eta_{\tilde{V}} = \frac{(n-1)}{2N + n/2}, & n \geq 2, \\ \epsilon_{\tilde{V}} = \frac{n}{4N + 2(n-1)}, \quad \eta_{\tilde{V}} = \frac{(n-1)}{2N + (n-1)}, & n < 2. \end{cases} \tag{56}$$

We set $\alpha = \gamma = 0$ to focus on discussing the contribution of the $R_{ab}T^{ab}$ term and obtain the field and potential function in Einstein's framework

$$\begin{aligned} d\chi &\approx \left[1 + 4\beta\kappa^2 \cdot \lambda M_{\mathrm{Pl}}^4 \left(\frac{\phi}{M_{\mathrm{Pl}}}\right)^n\right] d\phi \Rightarrow \chi \approx \phi + \frac{4\beta\lambda M_{\mathrm{Pl}}^{-n}}{n+1}\phi^{n+1}, \\ \phi &\approx \chi - \frac{4\beta\lambda M_{\mathrm{Pl}}^{-n}}{n+1}\chi^{n+1}, \end{aligned} \tag{57}$$

$$\tilde{V}(\chi) \approx \lambda M_{\mathrm{Pl}}^4 \left(\frac{\chi}{M_{\mathrm{Pl}}}\right)^n \cdot \left[1 + \frac{4(n+2)}{n+1}\beta\lambda\left(\frac{\chi}{M_{\mathrm{Pl}}}\right)^n\right]. \tag{58}$$

The slow-roll parameters with $\beta$ corrective items can then be exported

$$\epsilon_{\tilde{V}} = \frac{n^2}{2}\left(\frac{\widetilde{M_{\mathrm{Pl}}}}{\chi}\right)^2 + \frac{4n^2(n+2)}{n+1}\beta\lambda\left(\frac{\chi}{\widetilde{M_{\mathrm{Pl}}}}\right)^{n-2}, \tag{59}$$

$$\eta_{\tilde{V}} = n(n-1)\left(\frac{\widetilde{M_{\mathrm{Pl}}}}{\chi}\right)^2 + \frac{4n(n+2)(3n-1)}{n+1}\beta\lambda\left(\frac{\chi}{\widetilde{M_{\mathrm{Pl}}}}\right)^{n-2}, \tag{60}$$

$$\begin{aligned} N = \frac{1}{2n(n+1)}&\left[\left(\frac{\chi}{\widetilde{M_{\mathrm{Pl}}}}\right)^2\left(n + 1 - 8\beta\lambda\left(\frac{\chi}{\widetilde{M_{\mathrm{Pl}}}}\right)^n\right)\right. \\ &\left. - \left(\frac{\chi_{\mathrm{end}}}{\widetilde{M_{\mathrm{Pl}}}}\right)^2\left(n + 1 - 8\beta\lambda\left(\frac{\chi_{\mathrm{end}}}{\widetilde{M_{\mathrm{Pl}}}}\right)^n\right)\right]. \end{aligned} \tag{61}$$

All of the above approximations are preserved up to the first-order terms of $\beta$.

### 4.3. *Hilltop models*

The emergence of hilltop inflation was motivated in part by a predilection for concave potential function,[27] which can be written specifically as

$$V(\phi) = \Lambda^4\left[1 - \left(\frac{\phi}{\mu}\right)^p + \cdots\right], \tag{62}$$

where the ellipsis indicates the higher-order terms that make the potential function positive semi-definite. For the convenience of the calculation and the simplicity of the results, it is assumed that $\phi/\mu$ is small.

For the case of ignoring $\beta$ (we omit the higher-order term represented by the ellipsis in Eq. (62) in the following calculation),

$$\tilde{V}(\chi) = \frac{1+2\gamma}{(1+\alpha)^2}\Lambda^4\left[1-\left(\frac{\chi}{\tilde{\mu}}\right)^p\right], \quad \tilde{\mu} = \sqrt{\frac{1+\gamma}{1+\alpha}}\mu, \tag{63}$$

$$\epsilon_{\tilde{V}} \approx \frac{p^2}{2\kappa\chi^2}\left(\frac{\chi}{\mu}\right)^{2p}, \quad \eta_{\tilde{V}} \approx \frac{p(p-1)}{\kappa\chi^2}\left(\frac{\chi}{\mu}\right)^{2p}, \tag{64}$$

$$N \approx \begin{cases} \frac{\kappa\mu^2}{2p}\left\{\left[\left(\frac{\chi}{\mu}\right)^2 + \frac{2}{p-2}\left(\frac{\chi}{\mu}\right)^{2-p}\right] \right. \\ \qquad \left. -\left[\left(\frac{\chi_{\text{end}}}{\mu}\right)^2 + \frac{2}{p-2}\left(\frac{\chi_{\text{end}}}{\mu}\right)^{2-p}\right]\right\}, & p \neq 2, \\ \frac{\kappa\mu^2}{4}\left[\left(\frac{\chi}{\mu}\right)^2 - \left(\frac{\chi_{\text{end}}}{\mu}\right)^2 - 2\ln\left(\frac{\chi_{\text{end}}/\mu}{\chi/\mu}\right)\right], & p = 2. \end{cases} \tag{65}$$

When considering $\beta$ alone, $\alpha = \gamma = 0$ is set, and then

$$\begin{aligned} \mathrm{d}\chi &= \left[1 + 4\beta\kappa^2\Lambda^4\left(1-\left(\frac{\phi}{\mu}\right)^p\right)\right]\mathrm{d}\phi \Rightarrow \phi \\ &\approx \chi - 4\beta\kappa^2\Lambda^4\chi\left[1+\frac{1}{p+1}\left(\frac{\chi}{\mu}\right)^{p+1}\right], \end{aligned} \tag{66}$$

$$\tilde{V}(\chi) \approx \Lambda^4\left[1 + 8\beta\kappa^2\Lambda^4 + (4(p-4)\beta\kappa^2\Lambda^4 - 1)\left(\frac{\chi}{\mu}\right)^p\right], \tag{67}$$

$$\epsilon_{\tilde{V}} \approx \left[\frac{p^2}{2\kappa\mu^2} - \frac{4p^2(p-2)\beta\kappa\Lambda^4}{\mu^2}\right]\left(\frac{\chi}{\mu}\right)^{2p-2}, \tag{68}$$

$$\eta_{\tilde{V}} \approx \left[\frac{p(1-p)}{\kappa\mu^2} + \frac{4p(p-1)(p-2)\beta\kappa\Lambda^4}{\mu^2}\right]\left(\frac{\chi}{\mu}\right)^{p-2}, \tag{69}$$

$$\begin{aligned} N \approx \frac{\kappa}{p}&\left\{\left[\frac{\chi^2}{2} + \left(\frac{1}{p-2} + 4\beta\kappa^2\Lambda^4\right)\mu^2\left(\frac{\chi}{\mu}\right)^{2-p}\right]\right. \\ &\left. -\left[\frac{\chi_{\text{end}}^2}{2} + \left(\frac{1}{p-2} + 4\beta\kappa^2\Lambda^4\right)\mu^2\left(\frac{\chi_{\text{end}}}{\mu}\right)^{2-p}\right]\right\}, \quad p \neq 2. \end{aligned} \tag{70}$$

When $p = 2$, $f(R, T, R_{ab}T^{ab})$ does not generate a correction term for $N$ (up to the first-order). The results have been approximated by taking small amounts around zero for $\beta$ and $\phi/\mu$.

### 4.4. *Natural inflation*

Natural inflation means that the dynamical behavior of a scalar field is controlled by a periodic potential function[24–26]

$$V(\phi) = \Lambda^4\left[1 + \cos\left(\frac{\phi}{f}\right)\right], \tag{71}$$

whose behavior at the origin is similar to that of hilltop models, but differs on large scales. This potential function is usually reminiscent of spontaneous symmetry breaking and axis-like particles. $\Lambda$ and $f$ are dimensional parameters. It has been learned that this potential can motivate inflation when $\Lambda \sim M_{\rm GUT} \sim 10^{16}$ GeV and $f \sim M_{\rm Pl}$. However, in the GR framework it predicts a larger $r$ when $f \gtrsim 10\ M_{\rm Pl}$ and a smaller $n_{\rm s}$ when $f \sim M_{\rm Pl}$ compared with Planck observation. As before, we can calculate the slow-roll parameters when $\beta = 0$,

$$\epsilon_{\tilde{V}} = \frac{1}{2\kappa\tilde{f}^2}\left[\frac{\sin(\chi/\tilde{f})}{1+\cos(\chi/\tilde{f})}\right]^2, \quad \eta_{\tilde{V}} = -\frac{1}{\kappa\tilde{f}^2}\frac{\cos(\chi/\tilde{f})}{1+\cos(\chi/\tilde{f})}, \quad \tilde{f} \equiv \sqrt{\frac{1+\gamma}{1+\alpha}}f, \tag{72}$$

$$\epsilon_{\tilde{V}} - \eta_{\tilde{V}} = \frac{1}{2\kappa\tilde{f}^2}, \tag{73}$$

$$\exp\left[-\frac{N}{2\kappa\tilde{f}^2}\right] = \frac{\sin(\chi_{\rm end}/2\tilde{f})}{\sin(\chi/2\tilde{f})}. \tag{74}$$

The redefinition of the parameter $f$ can be seen as a re-engagement of the mass units:

$$\begin{aligned}\epsilon_{\tilde{V}}(\chi_{\rm end}) \equiv 1 \Rightarrow \chi_{\rm end} = 2\tilde{f}\arctan(\sqrt{2\kappa}\tilde{f}) \Rightarrow \sin\left(\frac{\chi}{2\tilde{f}}\right)\\ = \sqrt{\frac{2\kappa\tilde{f}^2}{1+2\kappa\tilde{f}^2}}\exp\left(-\frac{N}{2\kappa\tilde{f}^2}\right).\end{aligned} \tag{75}$$

Therefore, the slow-roll parameters can again be expressed in terms of the e-folding number $N$,

$$\epsilon_{\tilde{V}} = \frac{1}{{\rm e}^{N/\kappa\tilde{f}^2}(1+2\kappa\tilde{f}^2) - 2\kappa\tilde{f}^2}, \quad \eta_{\tilde{V}} = \frac{1}{{\rm e}^{N/\kappa\tilde{f}^2}(1+2\kappa\tilde{f}^2) - 2\kappa\tilde{f}^2} - \frac{1}{2\kappa\tilde{f}^2}. \tag{76}$$

Next, we reconsider the case $\alpha = \gamma = 0$ to study the nontrivial corrections related to $\beta$,

$$d\chi = d\phi\left[1 + 4\beta\kappa^2\Lambda^4\left(1+\cos\left(\frac{\phi}{f}\right)\right)\right] \Rightarrow \phi \approx \chi - 4\beta\kappa^2\Lambda^4\left[\chi + f\sin\left(\frac{\chi}{f}\right)\right], \tag{77}$$

$$\begin{aligned}\tilde{V}(\chi) &\approx \Lambda^4\left(1+\cos\left(\frac{\chi}{f}\right)\right)\left\{1 + 4\beta\kappa^2\Lambda^4\left[2 + 2\cos\left(\frac{\chi}{f}\right) + \frac{(\frac{\chi}{f} + \sin(\frac{\chi}{f}))\sin(\frac{\chi}{f})}{1+\cos(\frac{\chi}{f})}\right]\right\}\\ &\overset{\chi/f \ll 1}{=\!=\!=} \Lambda^4\left[2 - \frac{1}{2}\left(\frac{\chi}{f}\right)^2\right]\left\{1 + 4\beta\kappa^2\Lambda^4\left[4 - \left(\frac{\chi}{f}\right)^2\right]\right\},\end{aligned} \tag{78}$$

$$\epsilon_{\tilde{V}} \approx \frac{2}{\kappa f^2}\left[\frac{(\frac{\chi}{f})}{[4 - (\frac{\chi}{f})^2]^2} + \frac{8\beta\kappa^2\Lambda^4(\frac{\chi}{f})^2}{4 - (\frac{\chi}{f})^2}\right] \overset{\chi/f \ll 1}{=\!=\!=} \frac{(\chi/f)^2}{8\kappa f^2}(1 + 32\beta\kappa^2\Lambda^4), \tag{79}$$

$$\eta_{\tilde{V}} \approx -\frac{1}{\kappa f^2}\frac{2 + 8\beta\kappa^2\Lambda^4[4 - 5(\frac{\chi}{f})^2]}{4 - (\frac{\chi}{f})^2} \tag{80}$$

$$\xlongequal{\chi/f\ll 1} -\frac{2}{\kappa f^2}\left[4+\left(\frac{\chi}{f}\right)^2\right]-\frac{8\beta\kappa\Lambda^4}{f^2}\left[1-\left(\frac{\chi}{f}\right)^2\right]. \tag{81}$$

### 4.5. *Power-law inflation*

We consider the situation where the scaling function is power-law connected to cosmic time as our final example of applying our slow-roll equations to describe concrete models:

$$a(t)\propto t^p \Rightarrow H(t)=\frac{p}{t}. \tag{82}$$

At this stage, it is simple to calculate the slow-roll parameters defined by the Hubble parameter and its derivatives Eq. (38)

$$\epsilon=\eta=\frac{1}{p}. \tag{83}$$

For the case of $p>1$, $\epsilon=\eta<1$ is always established, so inflation can go on forever.

In standard GR, the variation of the scalar field with cosmic time and the potential function in this case are expressed in simple form

$$V(\phi)=V_0\exp\left(-\sqrt{\frac{2}{p}}\frac{\phi}{M_{\mathrm{Pl}}}\right),\quad \phi(t)=\sqrt{2p}M_{\mathrm{Pl}}\ln\left(\sqrt{\frac{V_0}{p(3p-1)}}\frac{t}{M_{\mathrm{Pl}}}\right). \tag{84}$$

For the current $f(R,T,R_{ab}T^{ab})$ situation, we can give the dependence of $V$ and $\phi$ on $t$, however, the explicit functional relationship between themselves is different to express in writing

$$V(\phi(t))=\frac{1}{\kappa}\frac{3(1+\alpha)p^2}{(1+2\gamma)t^2+12\beta\kappa p^2}, \tag{85}$$

$$\begin{aligned}\phi(t)=\ &\frac{1}{\kappa^{3/2}}\frac{(1+\alpha)\sqrt{1+2\gamma}}{6\sqrt{\beta(1+\gamma)}\gamma^2[(1+2\gamma)t^2+12\beta\kappa p^2]}\\ &\cdot\Big\{-6\gamma pt\sqrt{\beta\kappa(1+\gamma)(1+2\gamma)}+\sqrt{3}[(1+2\gamma)t^2+12\beta\kappa p^2]\\ &\cdot\left[\sqrt{1+\gamma}(2+\gamma)\operatorname{arccot}\left(\frac{2p\sqrt{3\beta\kappa}}{\sqrt{1+2\gamma}t}\right)-2\sqrt{1+2\gamma}\operatorname{arccot}\left(\frac{2p\sqrt{3\beta\kappa}}{\sqrt{1+\gamma}t}\right)\right]\Big\}.\end{aligned} \tag{86}$$

## 5. Inflationary Models: Numerical Results

In this section, we will show the numerical results corresponding to the first four potential models (here the last one is just mention as a toy model that can be solved analytically thoroughly). Before enumerating the $n_s$–$r$ curves, we can pre-make some judgments about numerical results, most of which come from analytical results accurate to the zeroth-order of $\beta$.

First, $\beta\neq 0$ is necessary to correct the trivial result of GR, which we have pointed out before, otherwise we can directly return to the action under GR by rescale fields and parameters. Second, in Eqs. (32a), (32b) and (32c), $\alpha$ and $\beta$ frequently appear in the form of parameter combination $\alpha - 4\beta\kappa^2 V(\phi)$, which means that the observables will have a very strong degeneracy between these two parameters. At the same time, there will also be a certain degeneracy between $\alpha$ and $\gamma$ and parameters of the same dimension as the scalar field, including $\mu$ for hilltop model and $f$ for natural model. Third, we can notice a natural constraint on parameters

$$\begin{cases} 1 + 2\gamma > 0, \\ (1+\gamma)(1+\alpha) > 0, \end{cases} \Rightarrow \begin{cases} \gamma > -1/2, \\ \alpha > -1. \end{cases} \tag{87}$$

The first inequality comes from the physics expectation that the sign of the potential function does not change after conversion to the Einstein framework; the second inequality comes from the purely mathematical meaning of the square root. Finally, since the potential function changes greatly in various models, this means that our numerical results are very sensitive to $\beta$; at the same time, $\alpha$ and $\gamma$ only mean to some extent It involves the redefinition of scalar fields and parameters, so it has little impact on the results.

Next, let us directly list and illustrate the figures of the observable measurements of $n_s$ and $r$. Here's how these scatter plots are drawn:

(i) By solving the cut-off conditions of the slow-roll approximation $\varepsilon_{\tilde{V}} = 1$ and $|\eta_{\tilde{V}}| = 1$ and taking the smaller value, which is used as the values of the scalar field at the end of slow-roll inflation $\phi = \phi_{\text{end}}$ and $\chi = \chi_{\text{end}}$. It should be noted that when the inflation field moves near this condition, the slow-roll approximation is actually no longer valid, and here we are satisfied with the approximate analysis.

(ii) According to the existing observation results, the value of e-folding number $N \sim 50$, and according to the convention of phenomenological analysis, we calculated the corresponding scalar fields $\phi$ and $N = 50$ and 60, respectively, as their values near the horizon.

(iii) Now we calculate the observables $n_s$ and $r$ corresponding to this scalar field value and plot them on the graph.

(iv) Finally, we change $\beta$ (and $\gamma$) and get a series of points, and compare it with *Planck* 2018 without/with BICEP/*Keck* result to study the impact of the correction term $R_{ab}T^{ab}$ on the predicted values. Due to degeneracy and insensitivity, we do not consider the case where the conformal rescaling constant $\alpha$ is nonzero.

The values of each parameter are marked above the picture. It needs to be explained that $\beta$ actually refers to the dimensionless parameter combination, for example $\beta \cdot \Lambda^4/M_{\text{Pl}}^4$ for Starobinsky inflation, hilltop model, natural inflation and $\beta \cdot \lambda$ for chaotic model. The unified legends are: ● for Planck TT, TE, EE + lowE + lensing with 95%($2\sigma$) CL, ■ for Planck TT, TE, EE + lowE + lensing with 68%($1\sigma$) CL, ◆ for +BK18 + BAO with 95%($2\sigma$) CL, ▲ for +BK18 + BAO with 68%($1\sigma$) CL, ▼ for theoretical predicted value with $N = 50$, ○ for theoretical predicted value with $N = 60$. This dataset of observations provided by the BICEP/*Keck* CMB experiment is open access.

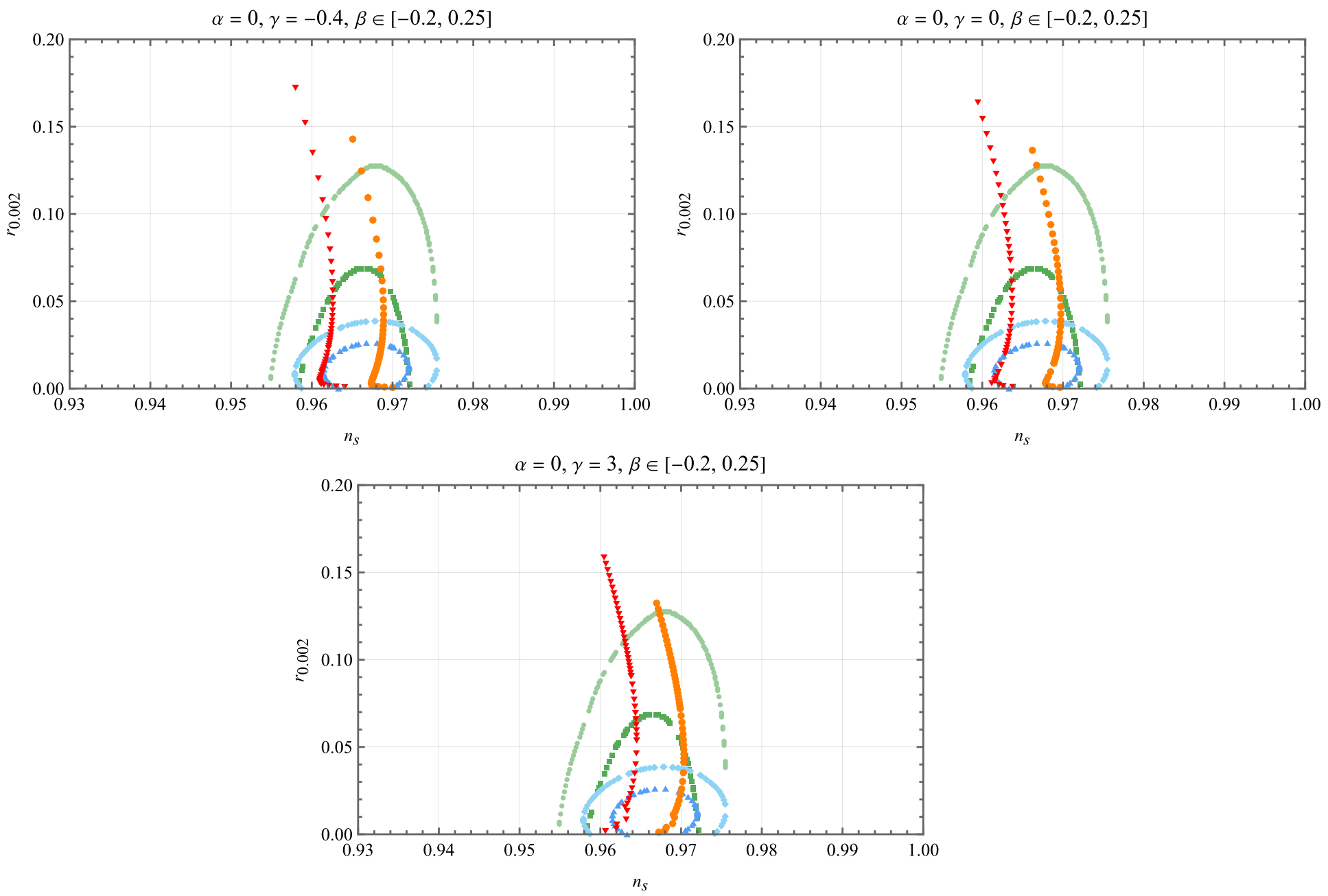

Fig. 2. Starobinsky inflation with different $\gamma$, $\beta$ goes from negative to positive as the scatter points go from bottom to top.

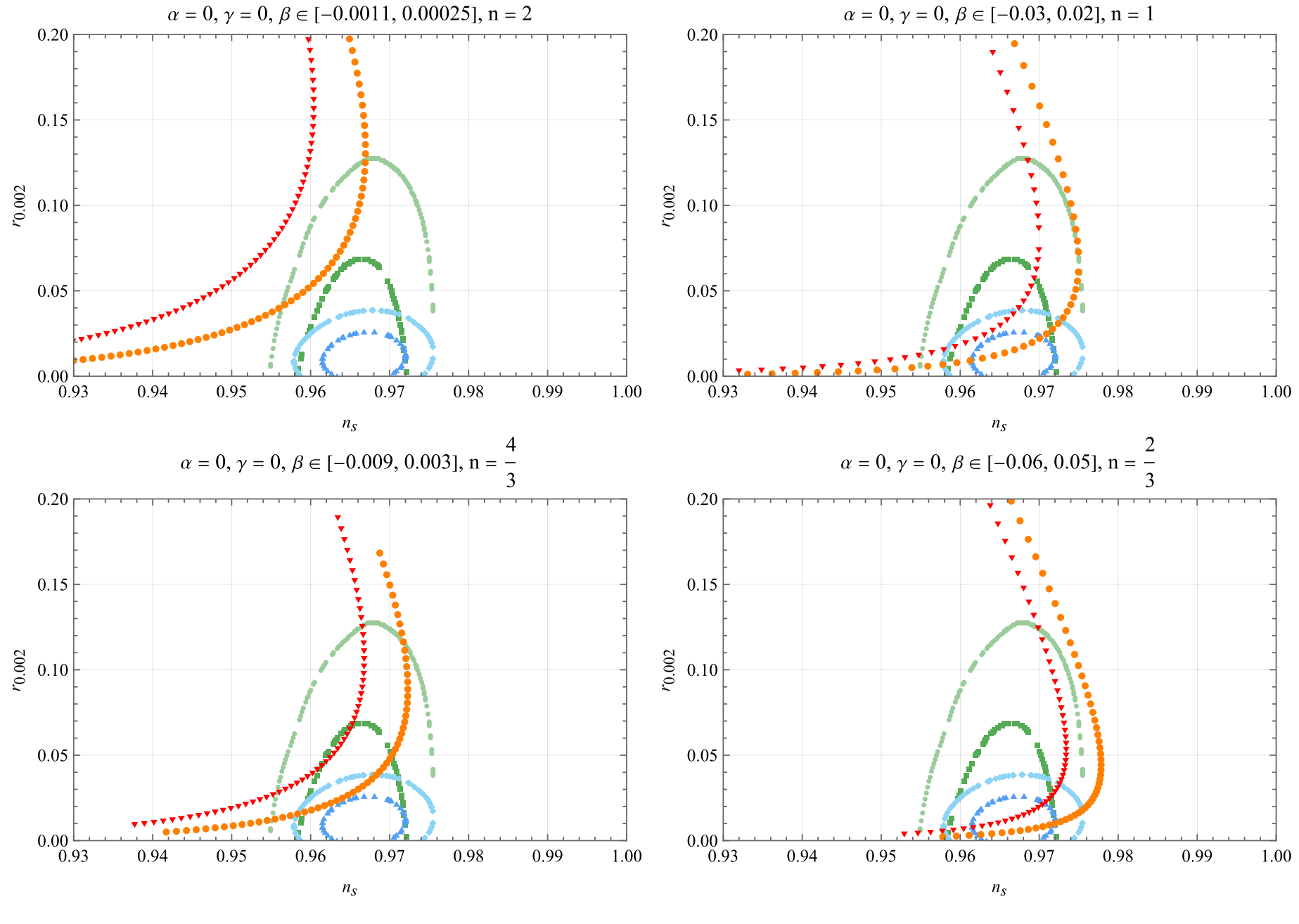

Fig. 3. Chaotic model with $\alpha = \gamma = 0$, $\beta$ goes from negative to positive as the scatter points go from bottom to top.

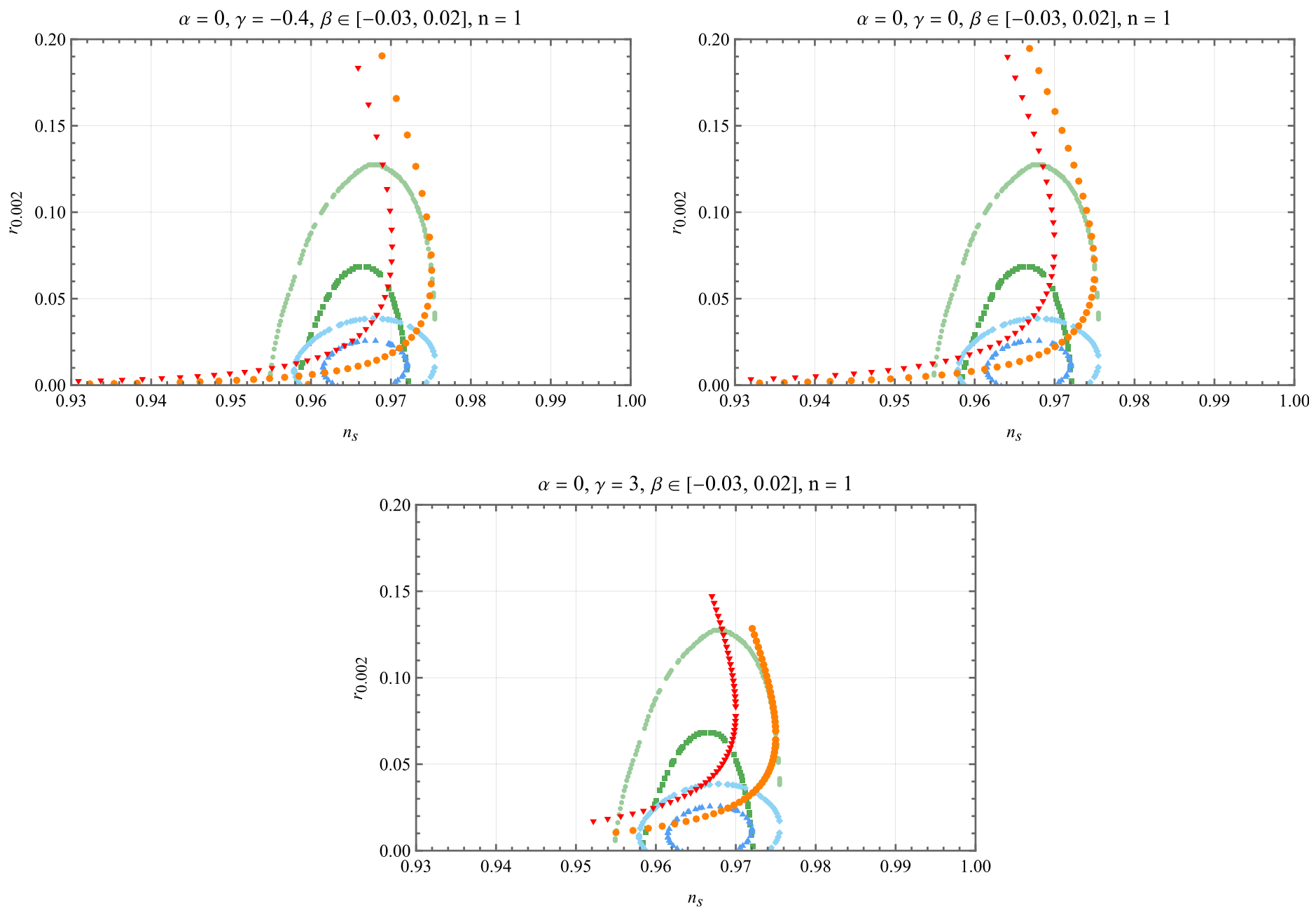

Fig. 4. Chaotic model with different $\gamma$, $\beta$ goes from negative to positive as the scatter points go from bottom to top.

Figure 2 shows the numerical results for Starobinsky inflation. The situation where $\beta < 0$ or $\beta > 0$ but not too large is accepted by the observation and changes in $\gamma$ do not seriously change this qualitative conclusion. In addition, the changing trends of the results corresponding to the value of $\beta$ on the positive and negative sides are obviously different, which comes from the asymmetry of the potential function reflected in Fig. 1.

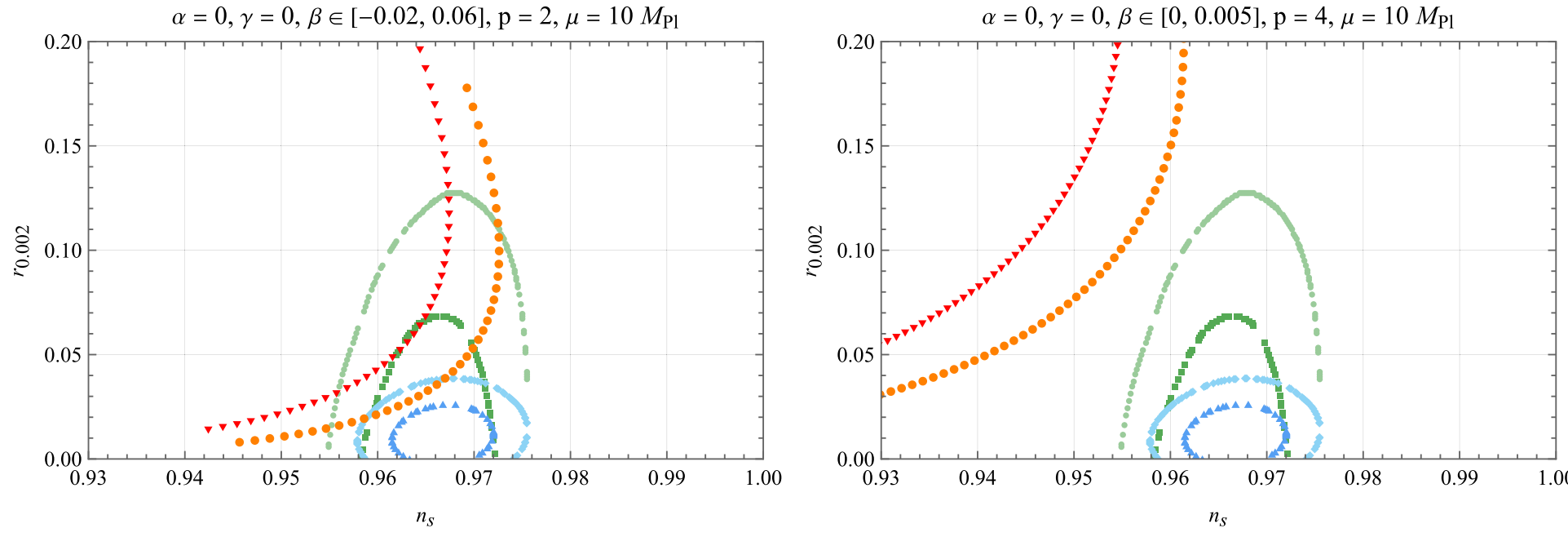

Fig. 5. Hilltop model with different $p$, $\beta$ goes from positive to negative as the scatter points go from bottom to top.

Figure 3 shows the numerical results for chaotic model with $n = 2$, $4/3$, $1$, $2/3$. Observations are very tolerant to this type of model, and theoretical predictions are consistent with them over a wide range of parameter space. Noting that $n = 1$ performs best among four sub-figures, we further studied the impact of changing $\gamma$ on the results at this time in Fig. 4. As we expected, the $n_s - r$ curve is not sensitive to $\gamma$, but it can still be observed that $\gamma < 0$ helps the curve to be closer to the result of BICEP/*Keck*.

Planck 2018 result gives parameter range $0.3 < lg(\mu/M_{\rm Pl}) < 4.85$ for hilltop quadratic model (hilltop model with $p = 2$) and $-2 < lg(\mu/M_{\rm Pl}) < 2$ for hilltop quartic model (hilltop model with $p = 4$). As a typical value, we take $\mu = 10$ and their corresponding predicted values are studied in Fig. 5. Obviously, $p = 2$ is closer to the observed value, while the curve of $p = 4$ is still far away from being consistent with the observation. Further, we compare the results of $p = 2$ with different $\gamma$ in Fig. 6.

Finally we look at the numerical results of natural inflation. The parameter constraint given by Planck 2018 result is $0.3 < lg(f/M_{\rm Pl}) < 2.5$. In studies of other literature, $f \gtrsim M_{\rm Pl}$ is usually taken. For our model, $f \sim M_{\rm Pl}$ gives poor predictive values, so we limit our analysis to $f \gg M_{\rm Pl}$, here taking typical values $f = 10M_{\rm Pl}$. When $\gamma = 0$, the predicted values do not agree well with the observed values. This situation is alleviated if we take $\gamma < 0$. The results in Ref. 22 show that it is possible to bring the predicted value into the observation area at this time, but due to the limitations of Eq. (87), we are more inclined to seek other ways to make the natural inflation consistent with the observation, such as Refs. 21, 22 and 38–42.

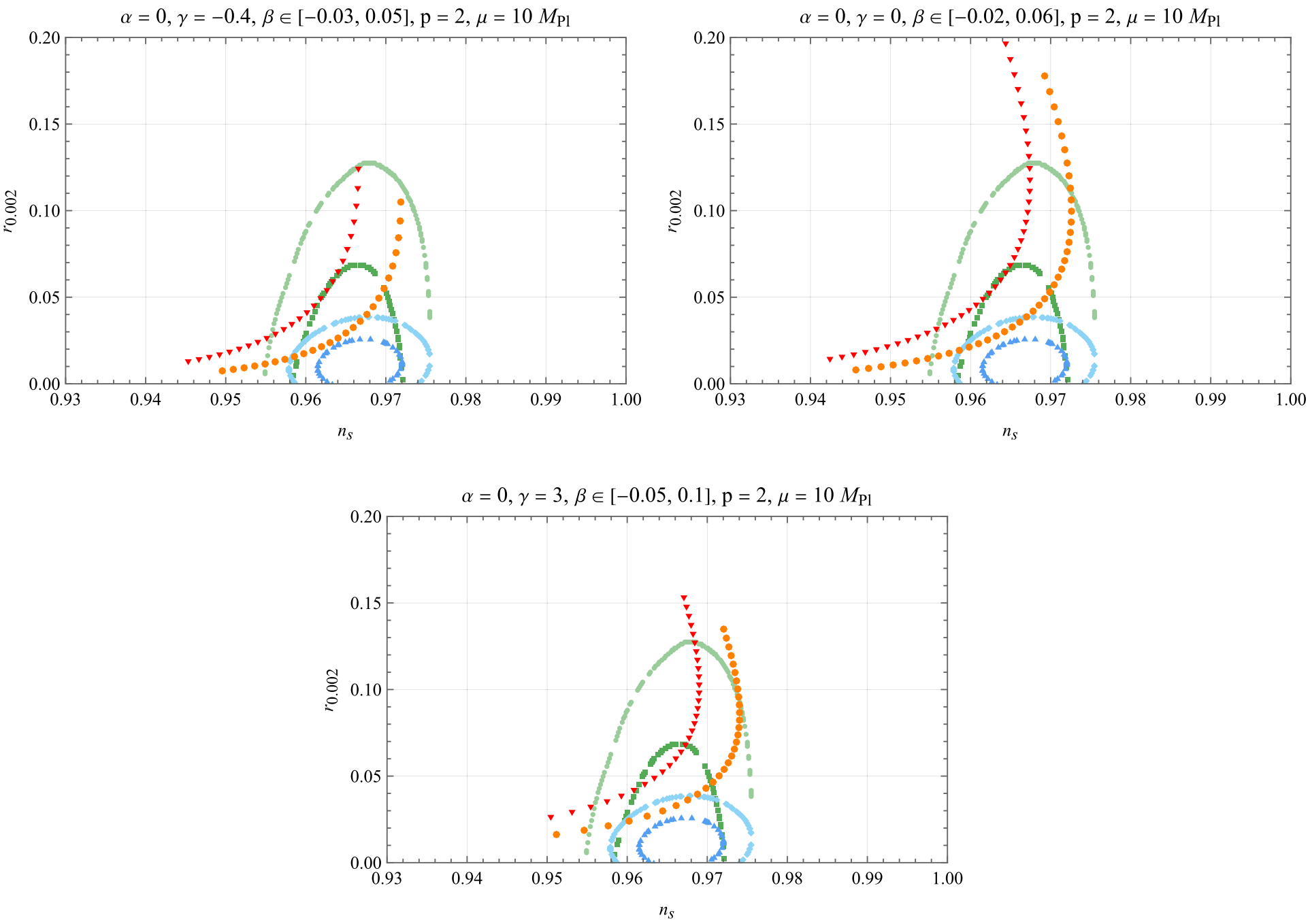

Fig. 6. Hilltop model with $p = 2$ (hilltop quadratic model) and different $\gamma$, $\beta$ goes from positive to negative as the scatter points go from bottom to top.

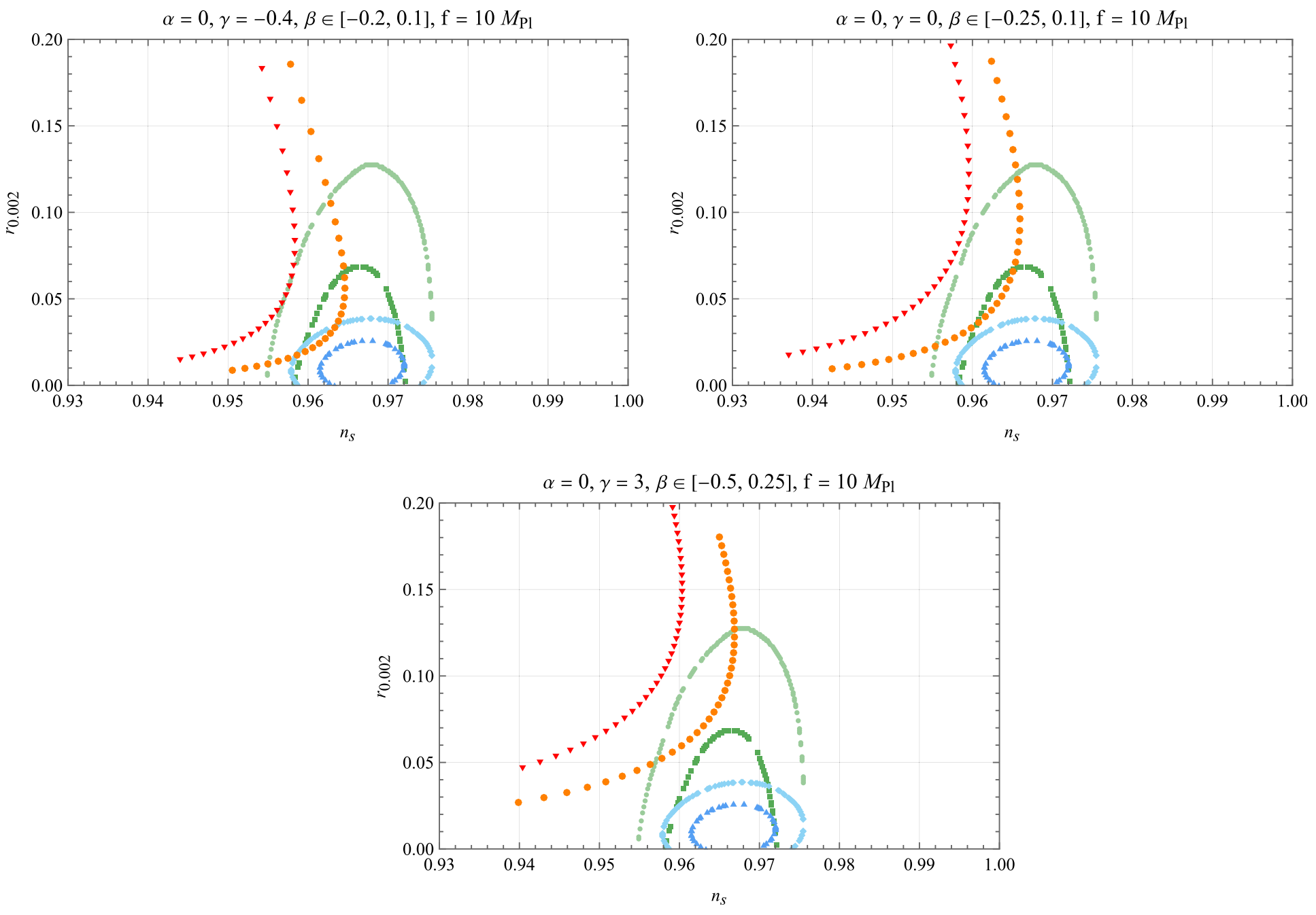

Fig. 7. Natural inflation with different $\gamma$, $\beta$ goes from negative to positive as the scatter points go from bottom to top.

Before concluding this section, it should be reminded again that the validity of the numerical results here strongly depends on the validity of the slow-roll approximation and the effectiveness of the Einstein framework under non-minimally coupled gravity theory. Therefore, the numerical results provided here are qualitatively correct. However further research is needed to provide quantitative corrections.

## 6. Summary

In this paper, we consider a modified theory of gravity with non-minimal matter-curvature coupling, namely $f(R, T, R_{ab}T^{ab})$, and study the dynamical behavior of cosmological inflation controlled by it. This kind of gravitational theoretical model has been widely used in recent years to study the deep coupling of spacetime and matter. We first reviewed the basic framework of $f(R, T, R_{ab}T^{ab})$ gravity, and derived the modified Einstein equation from the action. In the next step, we introduced the slow-roll approximation and obtained the field equation under it. We transform the field equations to the Einstein framework, thus introducing the slow-roll parameter. Afterwards, we focused on the analysis of different forms of potential functions and calculated their deviations from those in GR under the first-order approximation, including Starobinsky inflation, Chaotic models, Hilltop models, Natural inflation and Power-law inflation. Next, we display the numerical results corresponding to the above model in the form of scatter plots and compare them with the observed values of Planck 2018 data with/without correction. In this part, we analyze the changing trend of inflationary observables in parameter space.

In general, we believe that the various galactic and supergalactic structures in the Universe can be traced back to the earliest non-uniformity of matter density in the early Universe. Therefore, deviations from spatial homogeneity and isotropy are important for the study of the structure formation of the Universe. As an extension, we also derive the behavior of inflation under perturbation in appendix.

It is worth pointing out that although the action we initially analyzed did not contain a $RT$ mixing term, after taking the slow-roll approximation it is actually possible to include such a term. This finding extends the applicability of our work.

It should be noted that the analysis in this paper takes a large number of approximations in order to obtain analytical results. However, this makes the results strongly dependent on the validity of the slow-roll approximation taken for the gravitational field equations and the validity of the transformation to the Einstein framework. Near the end of the inflation, the slow parameter can no longer be neglected, which means that the higher-order terms in the equations without approximation will become significant.

The present work is not perfect for the study of slow-roll inflation under the $f(R, T, R_{ab}T^{ab})$ theory. There are still many unperformed analyzes that attract people, including rigorous and complete perturbation analysis, etc. We leave these topics as future work.

## Appendix A. Calculation Details

Nonzero components of Christoffel symbols:

$$\Gamma^0{}_{11} = \Gamma^0{}_{22} = \Gamma^0{}_{33} = a^2 H, \tag{A.1}$$

$$\Gamma^1{}_{01} = \Gamma^1{}_{10} = \Gamma^2{}_{02} = \Gamma^2{}_{20} = \Gamma^3{}_{03} = \Gamma^3{}_{30} = H. \tag{A.2}$$

Nonzero components of Ricci tensor and scalar:

$$R_{00} = -3\frac{\ddot{a}}{a} = -3(\dot{H} + H^2), \tag{A.3}$$

$$R_{11} = R_{22} = R_{33} = 2\dot{a}^2 + a\ddot{a} = a^2(\dot{H} + 3H^2), \tag{A.4}$$

$$R = 6\frac{\dot{a}^2 + a\ddot{a}}{a^2} = 6(\dot{H} + 2H^2). \tag{A.5}$$

Nonzero components of matter tensor:

$$T_{00} = \frac{1}{2}\dot{\phi}^2 + V(\phi), \quad T_{11} = T_{22} = T_{33} = a^2\left[\frac{1}{2}\dot{\phi}^2 - V(\phi)\right], \tag{A.6}$$

$$T \equiv g^{ab}T_{ab} = \dot{\phi}^2 - 4V(\phi), \tag{A.7}$$

$$\Theta_{00} = -\frac{3}{2}\dot{\phi}^2 - V(\phi), \quad \Theta_{11} = \Theta_{22} = \Theta_{33} = -a^2\left[\frac{1}{2}\dot{\phi}^2 - V(\phi)\right], \tag{A.8}$$

$$\Theta \equiv g^{ab}\Theta_{ab} = 4V(\phi), \tag{A.9}$$

$$R_{ab}T^{ab} = 3H^2\dot{\phi}^2 - 6(\dot{H} + 2H^2)V(\phi). \tag{A.10}$$

## Appendix B. Perturbations of Inflation Scalar Field

The inflation scalar field, we have discussed so far have good spatial translation invariance and isotropy. We attempt to discuss its behavior when it produces a perturbation that varies with time-space position right now. This section frequently appears in discussions of the structure formation of the universe. The perturbed field can be decomposed into a uniform background field and its variation

$$\phi(t, \vec{x}) = \bar{\phi}(t) + \delta\phi(t, \vec{x}). \tag{B.1}$$

Among them, $\bar{\phi}(t)$ satisfies Eq. (19) since it is equivalent to the background field we stated previously. The modified Klein–Gordon equation in covariant form, which directly results from the variation of the action equation (1) with respect to the scalar field, serves as our starting point:

$$(1 + \gamma + 2\beta\kappa R)g^{ab}\nabla_a\nabla_b\phi - 4\beta\kappa R^{ab}\nabla_a\nabla_b\phi - (1 + 2\gamma + 2\beta\kappa R)V'(\phi) = 0. \tag{B.2}$$

Substituting Eq. (B.1) into Eq. (B.2) considering the dynamic equation of the background scalar field, one can obtain

$$\begin{aligned}(1 + \gamma + 12\beta\kappa H^2)\ddot{\delta\phi} &+ (1 + \gamma + 12\beta\kappa H^2 + 8\beta\kappa\dot{H})3H\dot{\delta\phi} \\ &- (1 + \gamma + 12\beta\kappa H^2 + 8\beta\kappa\dot{H})a^{-2}\nabla_{\vec{x}}\delta\phi \\ &+ (1 + 2\gamma + 24\beta\kappa H^2 + 12\beta\kappa\dot{H})V''(\bar{\phi})\delta\phi = 0.\end{aligned} \tag{B.3}$$

Up to this point, the analysis was exact with the exception that we expanded the potential in a series while keeping the lower-order terms

$$V'(\bar{\phi} + \delta\phi) \approx V'(\bar{\phi}) + V''(\bar{\phi})\delta\phi. \tag{B.4}$$

After slow-roll approximation, we have

$$\begin{aligned}&(1 + \gamma + 12\beta\kappa H^2)\ddot{\delta\phi} + (1 + \gamma + 12\beta\kappa H^2)3H\dot{\delta\phi} \\ &\quad - (1 + \gamma + 12\beta\kappa H^2)a^{-2}\nabla_{\vec{x}}\delta\phi + (1 + 2\gamma + 24\beta\kappa H^2)V''(\bar{\phi})\delta\phi = 0.\end{aligned} \tag{B.5}$$

We can now expand $\delta\phi(t, \vec{x})$ into a Fourier series

$$\delta\phi(t, \vec{x}) = \sum_{\vec{k}} \delta\phi_{\vec{k}}(t)\mathrm{e}^{\mathrm{i}\vec{k}\cdot\vec{x}}. \tag{B.6}$$

In Fourier space, we have

$$(1+\gamma+12\beta\kappa H^2)(\delta\phi_{\vec{k}})^{\cdot\cdot}+(1+\gamma+12\beta\kappa H^2)3H(\delta\phi_{\vec{k}})^{\cdot} + \left[(1+\gamma+12\beta\kappa H^2)\left(\frac{k}{a}\right)^2+(1+2\gamma+24\beta\kappa H^2)V''(\bar{\phi})\right]\delta\phi_{\vec{k}}=0 \tag{B.7}$$

or

$$H^{-2}(\delta\phi_{\vec{k}})^{\cdot\cdot}+3H^{-1}(\delta\phi_{\vec{k}})^{\cdot}+\left[\left(\frac{k}{aH}\right)^2+\frac{1+2\gamma+24\beta\kappa H^2}{1+\gamma+12\beta\kappa H^2}\frac{m^2}{H^2}\right]\delta\phi_{\vec{k}}=0, \tag{B.8}$$

where $m^2\equiv V''(\bar{\phi})$.

If $H$ and $m^2$ change slowly during inflation, we make now an approximation where we treat them as constants. Further, by redefining $m$, Eq. (B.8) can be reduced to the form in GR

$$H^{-2}(\delta\phi_{\vec{k}})^{\cdot\cdot}+3H^{-1}(\delta\phi_{\vec{k}})^{\cdot}+\left[\left(\frac{k}{aH}\right)^2+\frac{\tilde{m}^2}{H^2}\right]\delta\phi_{\vec{k}}=0, \quad \tilde{m}=\sqrt{\frac{1+2\gamma+24\beta\kappa H^2}{1+\gamma+12\beta\kappa H^2}}m, \tag{B.9}$$

whose general solution is

$$\delta\phi_{\vec{k}}(t)=a^{-3/2}\left[A_{\vec{k}}J_{-\nu}\left(\frac{k}{aH}\right)+B_{\vec{k}}J_{\nu}\left(\frac{k}{aH}\right)\right], \quad \nu=\sqrt{\frac{9}{4}-\frac{m^2}{H^2}}, \quad a(t)\propto e^{Ht}, \tag{B.10}$$

where $A_{\vec{k}}$ and $B_{\vec{k}}$ are undetermined constants and $J$ is the Bessel function of the first kind.

If we ignore m/H for the cases where $m$ is small, i.e. $V(\phi)$ varies smoothly, the aforementioned formula degenerates to the scenario in GR. For small $\gamma$ and $\beta$, we keep to their first-order terms

$$H^{-2}(\delta\phi_{\vec{k}})^{\cdot\cdot}+3H^{-1}(\delta\phi_{\vec{k}})^{\cdot}+\left[\left(\frac{k}{aH}\right)^2+(1+\gamma+12\beta\kappa H^2)\frac{m^2}{H^2}\right]\delta\phi_{\vec{k}}=0. \tag{B.11}$$

We only analyzed the spatio-temporal perturbations of the inflation scalar field. Scalar and tensor perturbations of the metric tensor field itself are also of interest, but this content is beyond the scope of this paper.

## Acknowledgments

This paper is not driven by any profit organization, including any scientific projects or foundations. The vast majority of the calculation and plotting in this work were done using *Wolfram Mathematica* software. The author would like to highlight his gratitude to the developers of all the components of *xAct*,[43–46] which have greatly assisted him in his study and research of gravitational theory, including the present one. In addition, Ruoyang Cui from XJTU and Wenhao Dou from HEBUT provide guidance on plotting of numerical results, which is not related to physical research. The author is solely responsible for the correctness and reliability of the results in this paper.

## ORCID

Zhe Feng https://orcid.org/0000-0002-6377-8544